\documentclass[aps,nobibnotes,nofootinbib,preprint,superscriptaddress,]{revtex4}

\usepackage{amsmath}
\usepackage{latexsym}
\usepackage{graphicx}
\usepackage{subfig}
\usepackage{bm}        
\usepackage{amssymb}   
\usepackage{color}

\def\om{\omega}
\def\al{\alpha}
\def\be{\begin{eqnarray}}
\def\ee{\end{eqnarray}}
\def\beq{\begin{eqnarray}}
\def\eeq{\end{eqnarray}}

\def\k{{\bf k}}

\def\ba{\begin{eqnarray}}
\def\ea{\end{eqnarray}}
\def\beq{\begin{eqnarray}}
\def\eeq{\end{eqnarray}}

\def\mpl{M_{\rm Pl}}

\def\d{\mathrm{d}}
\def\p{{\cal P}}

\def\L*{{\cal L}_*}
\def\L{\mathcal{L}}
\def\({\left(}
\def\){\right)}

\def\nn{\nonumber}
\def\p{\partial}
\def\mn{_{\mu \nu}}
\def\stu{St\"uckelberg }
\def\p{\partial}

\def\<{\langle}
\def\>{\rangle}

\def\tK{\tilde{\cal K}}
\usepackage{amsmath}
\usepackage{amsfonts}
\usepackage{verbatim}
\usepackage{setspace}
\usepackage{url}
\usepackage{color}

\begin{document}
\title{On Cosmological Perturbations of Quasidilaton}
\preprint{UCSD/PTH 13-03}
\author{Guido D'Amico}
\author{Gregory Gabadadze}
\affiliation{Center for Cosmology and Particle Physics,
Department of Physics, New York University, New York,
NY, 10003, USA}
\author{Lam Hui}
\affiliation{Physics Department and Institute for Strings, Cosmology, and Astroparticle Physics,\protect \\ Columbia University, New York, NY 10027, USA }
\author{David Pirtskhalava}
\affiliation{Department of Physics, University of California, San Diego, \protect \\La Jolla, CA 92093 USA}

\vspace{0.2in}
\begin{abstract}
\vspace{0.2in}

A theory of the quasidilaton  is an extension of massive gravity by a scalar field, 
nonlinearly realizing a certain  new global symmetry of the Lagrangian.  It has been shown that unlike
pure massive gravity,  this theory  does admit homogeneous and isotropic  spatially flat solutions. 
Among the latter,  selfaccelerated solutions  attract a special attention. Previous studies of perturbations, 
performed in the decoupling limit, revealed one healthy scalar mode, while the second relevant 
scalar was not captured in that limit.  Here we study full cosmological perturbations above 
the simplest selfaccelerated background.  We show that the fluctuations  of
a mixed state of the quasidilaton and the helicity-0 graviton necessarily have a negative kinetic 
term at short distances, making this background unphysical.  In addition,  these cosmologies  
exhibit  an order one sensitivity to higher dimensional  terms suppressed by an energy 
scale that is parametrically higher than the strong coupling scale 
of the quasidilaton effective theory:  such terms include Galileons, Goldstone-like selfinteractions and 
derivatives of the quasidilaton coupled to curvature, none of which introduce extra Ostrogradsky 
states.  As  one consequence,  cosmology at the Hubble distances for this 
particular class of solutions depends on an unknown extension  of the quasidilaton 
below its strong coupling distance scale.  We note that non-FRW solutions that are similar to 
those of pure massive gravity should not necessarily suffer from these pathologies.

\end{abstract}

\maketitle
\section{Introduction and Summary}

The theoretical robustness and remarkable accuracy of General Relativity (GR) in describing gravitational interactions in a wide range of distance scales makes it one of the most successful physical theories of all times. Nevertheless, the late-time cosmic acceleration,  and the cosmological constant problems might be pointing towards a certain missing ingredient in the GR picture of gravity, calling for its modifications at distances of the order of the present-day Hubble scale. Massive gravity is perhaps the most conservative of such modifications. Whether or not an interacting graviton can have a nonzero mass without violating the theoretical consistency of GR has been a long standing problem originating from the work of Fierz and Pauli \cite{Fierz:1939ix}, who proposed a unique linear theory, propagating the five degrees of freedom of a massive spin-2 state. Generalizing the theory to the nonlinear level, however, has proven to be problematic, due to the emergence of an extra ghost-like state \cite{BoulwareDeser1972}; 
this 6th degree of freedom is referred to as the Boulware-Deser (BD) ghost. 

Recently, an explicit construction has been proposed in Ref. \cite{deRham:2010ik}, that 
gave an order-by-order Lagrangian for a single massive spin-2 state free of the BD ghost in a particular limit. 
In Ref. \cite{deRham:2010kj}  the Lagrangian  of 
\cite{deRham:2010ik} was resummed into a diffeomorphism invariant nonlinear theory, 
which was proposed as a ghost-free candidate for massive GR.  The presence of the hamiltonian constraint, necessary for projecting out the BD ghost was shown perturbatively, up to the quartic order in Ref. \cite{deRham:2010kj}. The 
full proof of the ghost-freedom to all orders  was given in  \cite{Hassan:2011hr}. Subsequently, the theory has also been shown to be BD ghost-free in the Lagrangian \cite{Mirbabayi:2011aa}, as well as in the vierbein formalisms  \cite{Hinterbichler:2012cn,Deffayet:2012zc}. 

An important property of the ghost-free theories of massive gravity is the existence of self-accelerated solutions with the Hubble scale of order of the graviton mass, as pointed out in Ref. \cite{deRham:2010tw}. 
Since then, these theories have been widely studied in the context of cosmology \cite{Koyama:2011xz, D'Amico:2011jj, Gumrukcuoglu:2011ew, Gumrukcuoglu:2011zh, Gratia:2012wt,Volkov:2012zb,DeFelice:2013awa,Volkov:2013roa}, spherically symmetric solutions and black holes \cite{Nieuwenhuizen:2011uq,  Koyama:2011yg, Chkareuli:2011te, Gruzinov:2011mm, Berezhiani:2011mt, Mirbabayi:2013sva}, as well as their quantum consistency \cite{deRham:2012ew, ArkaniHamed:2002sp}.
One interesting feature of cosmology in these theories is the absence of strictly homogeneous and isotropic spatially flat and closed backgrounds, while the obtained inhomogeneous solutions may still well approximate the observed world \cite{D'Amico:2011jj}. In particular, there are self-accelerated solutions for which the metric can be brought to the standard flat FRW form at the expense of having inhomogeneities in the \stu fields \cite{D'Amico:2011jj, Gratia:2012wt, Kobayashi:2012fz}. The spatially open FRW solutions have been found in \cite{Gumrukcuoglu:2011ew}. 
However, the above types of solutions generically suffer from nonlinear  instabilities\footnote{While the theories at hand are free of the BD ghost, one or more degrees of freedom out of the physical 5, can  flip the sign of their  kinetic terms on a nontrivial background, rendering it unstable. This is what we refer to as "ghost instability" here.} \cite{Gumrukcuoglu:2011zh, DeFelice:2012mx, D'Amico:2012pi}. 

Recently, it has been shown that an extension of massive GR by a scalar, the \textit{quasidilaton},  
nonlinearly realizing a certain 
new global symmetry, naturally reintroduces homogeneous and isotropic, 
spatially flat solutions, for certain values of the parameter characterizing its coupling to gravity \cite{D'Amico:2012zv}. Such solutions include selfaccelerated  backgrounds  that seem to be consistent with all immediate cosmological/astrophysical tests \cite{D'Amico:2012zv}. An important property, that automatically follows from the construction of the theory,  is its technical naturalness. Indeed, the form of the interactions of  the quasidilaton is protected by the new global symmetry of the theory. Moreover, in the decoupling limit, quasidilaton massive gravity (QMG) acquires yet another, enhanced global symmetry, reducing to a theory of two Galileons \cite{NicolisRattazziTrincherini2008} interacting with a tensor field \cite{deRham:2010ik}; all these terms  exhibit remarkable non-renormalization properties \cite{Luty:2003vm,Hinterbichler:2010xn,deRham:2012ew}. While the theory certainly stands out in this respect, it is crucial to address the question of stability of the spectrum of perturbations on such selfaccelerated backgrounds. 

The study of perturbations on the simplest selfaccelerated de Sitter (dS) solutions of QMG has been initiated  in \cite{D'Amico:2012zv}, in a certain high-energy (decoupling) limit of the theory. It has been shown however, that the decoupling limit analysis only captures one of the two scalar degrees of freedom; whether or not the second scalar propagates and is stable, remained an open question. In this Letter we perform  the full analysis,  and show that the second scalar is dynamical. However, we find that the kinetic term of this mode necessarily has a negative sign at short distances, rendering these simplest dS backgrounds unstable. 

We note, in addition, that these backgrounds exhibit  sensitivity to an unknown short distance physics:
the solution is fully modified by adding the Galileon and/or Goldstone-like  selfinteractions, as well as certain
derivative couplings to curvature tensors, which are consistent with all the symmetries of the Lagrangian, and do 
not introduce any Ostrogradsky ghost states. Some of these higher dimensional terms  are suppressed
by a scale that is parametrically higher that the strong coupling scale of QMG.  The order one sensitivity
to such terms makes the viability of these particular backgrounds questionable, even though the selfaccelerated 
solutions do generically exist even when these terms are included.  
In particular, even in the full quantum theory, where such terms are expected to be generated,  one should anticipate selfaccelerated backgrounds with the Hubble parameter $H$ of the order of the graviton mass; 
however,  as pointed out above, the precise nature and detailed properties of the spectrum of perturbations on these solutions are fully sensitive to an unknown extension of the quasidilaton below  its strong coupling distance scale.

While such a UV-sensitivity is a characteristic feature of the particular class of flat FRW solutions considered below and, as we will see, stems from the large rate of variation of the quasidilaton field on these solutions, there should exist  other solutions that closely resemble the inhomogeneous and/or anisotropic cosmologies of pure massive GR. In particular, for sufficiently large values of the parameter $\omega$, that controls the strength of the 
coupling of the quasidilaton to the rest of the fields, one expects the existence of inhomogeneous solutions that can recover the standard early cosmology to a great accuracy \cite{D'Amico:2011jj} . 

Moreover, even for the values of $\omega$ discussed  in this work, $0<\omega<6$, one may explore 
the existence of UV-insensitive inhomogeneous solutions, characterized by a small (or zero) expectation value of the quasidilaton, relying on the Vainshtein mechanism \cite{Vainshtein} in the
cosmological setup \cite{D'Amico:2011jj} for the other fields. This, however, is outside of the scope of the present work.

\section{Perturbations on the Self-Accelerated background}
\textbf{The framework:}
The theory we wish to consider is based on the recently formulated class of models of massive gravity, free of the Boulware-Deser ghost \cite{deRham:2010ik, deRham:2010kj}. We introduce  a special  
scalar $\sigma$ that gives rise to a certain new global symmetry of the Lagrangian.
The  symmetry transformation involves the scalar itself,  and the \stu 
fields $\phi^a,~a=0,1,2,3$, that are necessary if one wishes to work with a 
diffeomorphism-invariant action for massive GR.  
These four fields are scalars w.r.t. diffeomorphisms, but do transform under the 
Poincar\'e group of the internal space of $\phi^a$'s, as emphasized
in \cite{Siegel:1993sk, ArkaniHamed:2002sp}. The new global symmetry 
that we use as a building principle for the gravitational action involving
the scalar $\sigma$, is realized as follows: 
\be
\sigma \to \sigma - \alpha \, \mpl \, , \quad \phi^a \to e^{\alpha}\phi^a \,,
\label{newglob}
\ee
where $\alpha$ is an arbitrary symmetry transformation parameter. 
The rest of the fields in the Einstein frame\footnote{We define the Einstein frame in the standard way - 
the one for which the kinetic term for the graviton has the usual Einstein-Hilber form;  Jordan 
frame on the other hand will feature a kinetic mixing between the scalar $\sigma$ and the graviton.},  
and the physical coordinates $x^\mu$,  do not transform.
This symmetry 
fixes uniquely, \textit{modulo derivative interactions}, an extension 
of massive GR by the $\sigma$ field; in particular, one consequence of \eqref{newglob} is minimal coupling of matter to gravity in the Einstein frame (unlike Brans-Dicke theories for instance, which have matter coupled minimally to gravity in the Jordan frame).

The most general Lagrangian invariant under \eqref{newglob}, and excluding possible ghost-free derivative interactions\footnote{We will return to the derivative interactions in the last section.} of $\sigma$, is written in the Einstein frame as follows \cite{D'Amico:2012zv}
\beq
\label{eq:einstein}
\begin{split}
S_E &=  \int d^4 x~ \frac{\mpl^2}{2} \sqrt{-g}  \left[ R -\frac{\omega}{\mpl^2} g^{\mn}\p_\mu\sigma\p_\nu\sigma
- \frac{m^2}{4} \( \mathcal{U}_2(\tK) +\alpha_3 \mathcal{U}_3(\tK)+\alpha_4 \mathcal{U}_4(\tK) \) \right] \\ 
&+3\mpl ^2 m^2 \int d^4 x~\beta~ e^{4\sigma/\mpl}\sqrt{-det~ \p_\mu\phi^a\p_\nu\phi_a}+ \int \d^4 x  ~\sqrt{- g} \mathcal{L}_m(g_{\mn},\psi) \, .
\end{split}
\eeq
Here we have defined $\tK^\mu_{~\nu}=\delta^\mu_\nu-e^{\sigma/\mpl}\sqrt{g^{\mu\alpha}\p_\alpha\phi^a
\p_\nu\phi^b\eta_{ab}} \,$, where $\eta_{ab}$ is a matrix, that numerically coincides with the flat metric (we will use the mostly plus signature), and $\mathcal{U}_i$ are its specific antisymmetric polynomials 
\begin{subequations}
\begin{align}
\mathcal{U}_2&=2 \varepsilon_{\mu \alpha . .} \varepsilon^{\nu \beta . .} \tK^\mu_{\; \nu} \tK^\alpha_{\; \beta}\\
\mathcal{U}_3&=\varepsilon_{\mu \alpha \gamma .} 
\varepsilon^{\nu \beta \delta .} \tK^\mu_{\; \nu} \tK^\alpha_{\; \beta} \tK^\gamma_{\; \delta} \\
\mathcal{U}_4&=\varepsilon_{\mu \alpha \gamma \rho} \varepsilon^{\nu \beta \delta \sigma} \tK^\mu_{\; \nu} \tK^\alpha_{\; \beta} \tK^\gamma_{\; \delta} \tK^\rho_{\; \sigma} \, ~.
\label{potential}
\end{align}
\end{subequations}
The dimensionless constants $\alpha_{3,4}, \beta,~\omega$, represent the four free parameters of the theory (the mass is another free parameter, but for phenomenological reasons, we set it to be of order of the present Hubble scale.) We will refer to the first two terms of \eqref{eq:einstein} as the 'gravity sector', while the last one specifies the coupling of matter fields $\psi$ to it, which, in order to respect the symmetry \eqref{newglob}, is given by the regular general-relativistic minimal coupling to the Einstein frame metric $g_{\mn}$. As emphasized above, the defining property of the gravity sector is its invariance under the global one-parameter group \eqref{newglob}, which is related to (nonlinearly realized) scale transformations and is explicitly broken by couplings to matter.
\vskip 0.5cm
\textbf{Background cosmology:}
The theory at hand has been extensively studied in Ref. \citep{D'Amico:2012zv}. In particular, it has been shown to be BD ghost-free at the full non-linear level; moreover, it has been found that in the decoupling limit it reduces to a bi-galileon theory, thus inheriting all of the properties, characteristic of the galileon field theories \cite{NicolisRattazziTrincherini2008}. This includes ghost-freedom, a successful implementation of the Vainshtein screening, non-renormalization, etc. (see also \cite{Creminelli:2010ba, Creminelli:2012my} for models of the early universe, based on galileons.) Most importantly for the present work however, quasidilaton extended massive gravity has been shown to admit strictly homogeneous and isotropic flat FRW cosmologies - in sharp contrast to the pure massive gravity, as well as to its previously studied extensions. The absence of the standard cosmologies in the latter theories is directly linked to ghost-freedom. Indeed, the same constraint that removes the BD ghost from the spectrum, severely constrains the homogeneous and isotropic evolution of the universe, essentially restricting the scale factor of such a universe to be static, $\dot a=0$. Standard FRW evolution however becomes possible  in QMG, due to the presence of an extra scalar field, whose cosmological variation allows to lift the above restriction. We briefly describe how this works in the appendix, concentrating on the homogeneous and isotropic field configurations in the Einstein frame theory and considering the asymptotic dynamics, when the matter/radiation has redshifted away\footnote{We refer the reader to  \cite{D'Amico:2012zv} for more details.}. In particular, in terms of the three free parameters of the theory (we assume the graviton mass is of order of the present Hubble scale), the Hubble constant is given by
\be
\label{eq:frwc}
\begin{split}
H^2 =
~\frac{m^2 \bigg( \frac{1}{4} (\al_3 + 4 \al_4) c^3 
- (1+\frac{3}{2} \al_3 + 3 \al_4) c^2 
+ (3 + \frac{9}{4} \al_3 + 3 \al_4) c
- (2 + \al_3 + \al_4) \bigg) }{1-\frac{\omega}{•6}}~,
\end{split}
\ee
where $c$ denotes a specific combination of the three parameters $\alpha_{3,4}$ and $\beta$, which solves the quartic equation, given in \eqref{sols}. For the particular case of $\beta=0$, the two interesting solutions are given by
\beq
\label{cs1}
 c = \frac{3 \alpha_3 + 8 \alpha_4 \pm \sqrt{9 \alpha_3^2 - 64 \alpha_4}}{8 \alpha_4}~.
\eeq
The de Sitter solutions exist in the latter case as far as the following conditions are met \cite{D'Amico:2012zv}
\be
\label{con32}
\alpha_3\neq 0, \qquad 0<\alpha_4<\frac{\alpha^2_3}{8}, \qquad 0\leq \omega< 6 ~.
\ee
The last condition, $0\leq \omega< 6$, is necessary for the well-definiteness of the solution at hand, regardless of whether $\beta$ is zero or not.
\vskip 0.5cm
\textbf{Perturbations:}
The study of perturbations on the above self-accelerated background has been initiated in \cite{D'Amico:2012zv}, where it has been shown that the conditions for the existence of the dS solutions also guarantee that ghost instabilities are absent for both the tensor and vector perturbations (see also \cite{Blas:2009my} for a discussion of stability in a general Lorentz-breaking massive gravity on cosmological backgrounds.) The remaining open question concerns the ghost-freedom of the scalar sector. 

As we will see shortly, the situation is complicated by an intricate nature of the scalar perturbations. There can be up to two propagating scalar degrees of freedom on a generic background (the helicity-zero graviton and the quasidilaton). In massive gravity and its closely related theories however, cosmological backgrounds commonly happen to be degenerate, with some degrees of freedom losing their dynamics, or becoming infinitely strongly coupled. In QMG, one can show that the decoupling limit treatment only captures one mode \cite{D'Amico:2012zv}, without being able to say anything about the other; one therefore has to resort to the full treatment of the scalar perturbation Lagrangian, which we perform in the rest of the present section. The resulting analysis is rather technical. We will suppress lengthy expressions whenever possible, providing instead a detailed account of the procedure used and results obtained. 

Our starting point is the perturbation Lagrangian given in \eqref{sublambdatilde0} - \eqref{gammas} in the unitary-like gauge. Various conventions and definitions used are given in the appendix; in particular, we will work with the conformal time $\tau$, with a prime denoting differentiation with respect to it. The action, charecterized by the parameters $\gamma_i$ explicitly given in \eqref{gammas}, involves the perturbation of the quasidilaton $\zeta$, coupled to the metric perturbations $h_{00}$, $h_{0i}$ and $h_{ij}$.

It is relatively easy to deal with the tensor and vector modes (see \cite{Blas:2009my, D'Amico:2012zv}.) In particular, one can show, that as far as $\gamma_5>0$, there are no ghost instabilities in either of the two sectors. For the particular case of $\beta=0$, this always holds as far as the dS existence conditions \eqref{con32} are satisfied. Gradient instabilities on the other hand never occur in the tensor sector, while their absence for the vector modes requires
\beq
\gamma_6<0~,
\eeq
which is also the condition for the absence of tachyonic instability for the tensor modes.

Turning to the discussion of the scalar sector, we present the essential steps of the calculation, along with the basic results.  

With the following decomposition
\beq
\label{dec}
h_{00}=\psi, \qquad h_{0i}=\p_i v, \qquad h_{ij}=\frac{\p_i\p_j}{\Delta}\rho+
\delta_{ij}\phi~,
\eeq
the Lagrangian for scalar perturbations in the spatial momentum space\footnote{To avoid complicating the presentation, we will not change the notation for the spatially Fourier-transformed fields.} can be written as follows (to simplify expressions, we will make use of the following relations, $\gamma_6=-\gamma_7$, $\gamma_4=-\gamma_1/6$, that hold for the most general choice of the parameters)
\ba
\mathcal{L}&=&\frac{a^2}{2}\bigg [-\frac{3}{2}(\phi'+a H\psi)^2-\frac{1}{2} \k^2\phi(2\psi-\phi)-(\phi'+a H\psi)(2\k^2 v+\rho')+\omega(\zeta'^2-\k^2\zeta^2)\nn \\ &+&\omega a H(\psi+3\phi+\rho) \zeta' -2\omega a H \k^2 v \zeta+a^2 \gamma_1 \psi\zeta+a^2  \gamma_2 (3\phi+\rho) \zeta + a^2\gamma_3 \psi^2+ a^2\gamma_4 (3\phi+\rho)\psi\nn \\ &+&a^2\gamma_5\k^2 v^2- 2 a^2 \gamma_6 (3\phi^2+2\phi\rho)+a^2 \gamma_8\zeta^2 \bigg ]~.\nn
\ea
An immediate observation is that $\psi$ and $v$ enter without time derivatives. Integrating out these fields, one obtains a fairly complicated effective Lagrangian $\mathcal{L}_3$ for $\rho$, $\phi$ and $\zeta$, the exact form of which we will not reproduce here. An important property of this Lagrangian however, is that the two-derivative kinetic matrix $K_3$, defined for the fields $\Phi=(\rho,\phi,\zeta)^T$ by
\beq
\mathcal{L}_3= \Phi'^T K_3 \Phi'+\dots~,
\eeq
is degenerate, having one zero eigenvalue with the corresponding eigenstate being $\rho_1\equiv 2\phi+\zeta$. This combination is thus non-dynamical and will have to be integrated out in order to obtain the effective action for the two remaining dynamical states. Of course, the vanishing of the determinant for the kinetic matrix $K_3$ should be no surprise (and is a nice consistency check of the computation), since the BD ghost-freedom of the theory guarantees that there are no more than two scalar degrees of freedom propagating on an arbitrary background. We take the latter two to be $$\rho_2\equiv -\frac{1}{2}\phi'+\zeta', \qquad \rho_3\equiv \rho~.$$ Solving the $\rho_1$ - equation of motion and substituting the solution back into the action yields the final Lagrangian $\mathcal{L}_2$, that determines the dynamics of the two propagating scalars. In the short-wavelenth limit, ($|\k|\gg m$), the kinetic matrix $K_2$ for the fields $\rho_2$ and $\rho_3$ drastically simplifies: it is degenerate, with the non-zero eigenvalue being $Z^{UV}_2=a^2\omega /2$ for the eigenstate $\rho_2$, while the kinetic term for $\rho_3$ vanishes in this limit\footnote{This explains the absence of the second scalar mode in the decoupling limit, as found in \citep{D'Amico:2012zv}.}. In order to capture the dynamics of $\rho_3$ therefore, we have to go to higher order in the $|\k|^{-1}$ expansion. The easiest way to do this is to deal with the determinant of the kinetic matrix, which has the following expansion 
\beq
\text{det}~ K_2=-\frac{•a^8(6-\omega)\gamma_1^2}{96}\frac{1}{\k^4}+\mathcal{O}\(\frac{1}{\k^6}\)~.
\eeq
In order to obtain the latter equation, the following relations have been used
\beq
\label{rel}
\theta_1\equiv-\gamma_1^2+6\gamma_1\gamma_5+9 \omega H^2\gamma_5=0, \qquad \theta_2\equiv 6\gamma_2+24\gamma_7+\gamma_8-9\omega H^2=0~,
\eeq
which can be obtained on the basis of the constraint equation \eqref{sols}, and drastically simplify the expressions at hand.
The leading contribution to the second eigenvalue of the kinetic matrix in the UV is therefore $$Z^{UV}_3=\frac{a^6 (\omega-6)\gamma_1^2}{48\omega \k^4}~.$$ 
Unfortunately, the condition for the existence of a well-defined de Sitter solution, $\omega<6$, constrains this quantity to be negative, thus signalling a ghost instability for the short wavelength modes. The full expression for the determinant, using the relations \eqref{rel}, can also be written in a simple form
\beq
\text{det}~ K_2=\frac{a^8 (6-\omega) \gamma_1^2}{24\k^2\big [a^2\omega(6-\omega)H^2-4\k^2\big]}~.
\eeq
One can see that in going from the UV to the IR, the determinant changes sign exactly once, being manifestly positive for the long-wavelength modes. 

\vskip 0.5cm

\textbf{UV dispersion relations:} As we will see momentarily, the vanishing of the $1/\k^2$ piece in the expansion of the determinant (proportional to $\theta_1$) is not accidental: it is required in order to obtain the regular dispersion relations $E^2\sim \k^2$  for the short wavelength modes\footnote{Of course, the fact that the leading UV contribution to the kinetic term of the $\rho_3$ field goes as $\k^{-4}$ stems from the specific normalization of the metric perturbation $\sigma$ in \eqref{dec}.}. The kinetic part of the Lagrangian in the $\k\gg m$ limit can be written as follows:
\beq
\mathcal{L}^{UV}_2\supset \frac{a^2\omega }{2}\tilde\rho_2'^{~2}+\frac{a^6 (\omega-6)\gamma_1^2}{48\omega \k^4}\rho_3'^{~2}+\dots~,
\eeq
where ellipses denote possible corrections of higher order in the short-distance expansion. There is a correction to one of the eigenstates of the kinetic matrix, 
\beq
\tilde\rho_2= \rho_2+\frac{\gamma_1 (\omega-6) a^2}{12 \k^2\omega}~\rho_3~,
\eeq
which is retained above, since it gives an $\mathcal{O}(1)$ effect in the UV, as will become clear below (this can easily be seen by canonically normalizing the $\rho_3$ field.) 

As a next step, we have a look at the one-derivative part of the action. There are four possible types of terms: $\rho_2\rho_2'$, $\rho_3\rho_3'$, $\rho_2\rho_3'$ and $\rho_2'\rho_3$. The first two of these are total derivatives and, upon integration by parts, can be reduced to non-derivative mass/gradient terms (we should keep in mind  that these operators come with coefficients that depend on the conformal time through $a(\tau)$, thus giving nonzero contributions.) It is convenient to arrange the rest of the one-derivative terms into combinations $\rho_2\rho_3'\pm \rho_2'\rho_3$. The first of these is again a total derivative and therefore only contributes to the non-derivative part, while the hermitian antisymmetric combination starts from the order $1/\k^2$ and enters into the Lagrangian as follows \footnote{There is a seemingly accidental cancellation in that the piece of order $\k^0$ only enters with the total derivative combination $\rho_2\rho_3'+ \rho_2'\rho_3$, but not with the orthogonal one. Again, this has to be the case if the dispersion relation is to be quadratic in momenta in the UV.}
\beq
\mathcal{L}^{UV}_2\supset \frac{c_1}{\k^2} \(\rho_2\rho_3' - \rho_2'\rho_3\)~.
\eeq
At the level of zero time derivatives, there are three types of terms: $\rho_2^2$, $\rho_2\rho_3$ and $\rho_3^2$. Keeping track of all partial integrations done up to this stage, and leaving only the leading contributions in the short-distance expansion, the relevant terms are given as follows
\beq
\mathcal{L}^{UV}_2\supset c_{01}\k^2\rho_2^2+c_{02}\rho_2\rho_3+\frac{c_{03}}{\k^2}\rho_3^2 + \dots~.
\eeq
In order to extract the physical scalar spectrum on the given de Sitter background, we will have to work in terms of the eigenstates of the kinetic matrix $\tilde\rho_2$ and $\rho_3$, canonically normalized\footnote{Note, that the normalization factors $Z$ explicitly depend on time through the scale factor. Canonical normalization will thereforre produce extra one and zero-derivative terms. These will however be subleading at short distances.} as follows, $\tilde\rho_{2}\to \tilde\rho_{2}/\sqrt{Z^{UV}_2}$, $\rho_3\to \rho_3/\sqrt{Z^{UV}_3}$. This procedure puts the kinetic term in the form, invariant under orthogonal transformations of the two dynamical fields, that one can use to diagonalize the mass/gradient matrix. For perturbations of frequency $E$, exceeding the characteristic time of the Hubble expansion, it is convenient to Fourier transform in time as well, $'\sim E$. It is easy to see now that the kinetic terms are of order $E^2$ at high energies, one-derivative terms are of order $m E$, while the gradient terms are $\sim \k^2$. This shows that the dispersion relations are of the regular $E\sim \k$ form in the UV, so that the one-derivative contribution can be completely neglected. 

\vskip 0.5cm
\section{UV Sensitivity}

We have shown above, that the scalar perturbation spectrum on the homogeneous and isotropic de Sitter solutions obtained in quasidilaton extended massive gravity in \cite{D'Amico:2012zv}, for the most general choice of the free parameters, features all dynamical modes present in the theory. Moreover, one combination of scalars is necessarily a ghost with the regular dispersion relation $E^2\sim\k^2$ at short distances, rendering these backgrounds unstable. 

However, the theory considered above is not the most general one, consistent with the global symmetry \eqref{newglob}. Indeed, while we have certainly included all possible \textit{non-derivative} interactions of the quasidilaton field, there still are ghost-free derivative interactions that may be added to the quasidilaton Lagrangian. 
For instance, one could add the covariantized  cubic Galileon (see \cite {NicolisRattazziTrincherini2008} for 
general Galileons, and \cite {Deffayet:2009wt} for their covariantization)
\beq
\sqrt{g}\, { g^{\mu\nu}{\nabla_\mu \sigma}{\nabla_\nu\sigma} \, g^{\alpha\beta}\nabla_\alpha \nabla_\beta \sigma \over 
\Lambda_3^3}\,.
\label{gal3}
\eeq
This term is invariant w.r.t. shifts of $\sigma$ by a constant, and hence preserves the symmetry \eqref{newglob}.
Moreover,  on an arbitratry background it gives rise to the second order equation of motion. 
Note that here we chose the suppression scale of this  operator to be 
$\Lambda_3^3 = M_{\rm Pl}m^2$, since this is the strong coupling scale of the quasidilaton 
theory on the Minkowski space \cite{D'Amico:2012zv}.  At this scale the theory needs a
UV  extension, and adding such a term would in general be motivated in the full quantum theory.  

Can this and other Galileon  terms  affect the conclusions of our previous section? At first sight the answer
 seems negative, since the selfaccelerated solution we are discussing is characterized by the curvature scales of order $1/m$, that are 
 much greater than $\Lambda_3^{-1}$. However, this  naive expectation is questionable: 
 for the homogeneous and isotropic cosmology  with these new terms  we still have the
 relation $${\dot \sigma} \sim M_{\rm Pl}H,$$
  which is not modified by 
 the inclusion of the Galileons. Due to this scaling,  the effects of the Galileon terms 
 cannot in general be ignored in our considerations.
 
 Furthermore, along the same lines,  one could add to the quasidilaton Lagrangian 
the symmetry-preserving Goldstone-like self-interactions of the $\sigma$ field:
\beq
\label{goldst}
\sqrt{g} \, \frac{( g^{\mu\nu} \nabla_\mu \sigma \nabla_\nu \sigma )^{n}}{\Lambda_2^{4n-4}},
\eeq
where the natural scale for these operators is $\Lambda_2\equiv (\mpl m)^{1/2}$. 
Note that $\Lambda_2 >>\Lambda_3$. Therefore, these operators are 
suppressed by the scale that is higher than the flat-space strong coupling scale 
of QMG.  Naively, such  terms should be irrelevant at scales $1/m$.
Nevertheless, due to the relation, ${\dot \sigma} \sim M_{\rm Pl}H  \sim \Lambda_2^2$,
these  terms do modify the selfaccelerated background as well as properties of perturbations on it.
To see this more explicitly,  let us for simplicity concentrate on the term  with $n=2$. 
Since it does not include the St\"uckelberg scalars, the constraint equation in \eqref{sols} is not modified,
giving ${\dot \sigma \sim M_{\rm Pl}H}$, while the Friedmann equation receives a correction 
proportional to $H^4$. Restoring all the scales,   the Friedmann equation takes  the following schematic form:
\beq
y_1 H^2+y_2\frac{H^4}{m^2}=y_3 m^2~,
\eeq
where the coefficients $y_i$ are some functions of the dimensionless parameters of the theory. One can therefore anticipate the de Sitter solutions with $H\sim m$, just as in the theory without the derivative $\sigma$
self-interactions. Moreover, the structure of perturbations on this solution will also get modified, 
since the terms  \eqref{goldst} generically contribute to all pieces in the quadratic perturbation Lagrangian \eqref{pert_h}. 

Moreover, even if the tree-level theory is taken to be the one considered above, quantum loops should generate the operators of the form \eqref{goldst}. Naively, the magnitude of these operators in \eqref{goldst} seems to be smaller than what one would expect from knowing that the cutoff of the theory is $\Lambda_3$. However, the fact that they should be suppressed by a scale higher than $\Lambda_3$, stems from the special nature of the Minkowski space decoupling limit of the theory. Indeed, as noted in \cite{D'Amico:2012zv}, in the decoupling limit the theory acquires an enhanced galilean symmetry, under which the quasidilaton shifts as
\beq
\sigma\to \sigma+b_\mu x^\mu~.
\eeq
Goldstone-type interactions are not invariant under this transformation, and would not be generated in the decoupling limit, if absent from the tree-level theory. This means that whatever the correction is in the full theory, it  should not survive taking the decoupling limit, implying that $\Lambda_2$ sould be taken as the suppressions scale for this operator.

The Galileon and Goldstone-like  interactions do not exhaust the full list of all possible ghost-free extensions of the model. Indeed, there are certain derivative couplings to the curvature tensors \cite{Chkareuli:2011te,deRham:2011by}, that represent a covariantization of the ghost-free scalar tensor interactions of the decoupling limit of massive GR \cite{deRham:2010ik}. In particular, one can extend the Lagrangian by the following interaction
\beq
\mpl^2 m^2~\sqrt{g} ~\frac{G_{\mn}\nabla^\mu\sigma\nabla^\nu\sigma}{\Lambda_3^6}~,
\eeq
as well as a higher-order term 
\beq
\mpl^2 m^2~\frac{L_{\mu\alpha\nu\beta}\nabla^\mu\sigma\nabla^\nu\sigma\nabla^\alpha\nabla^\beta\sigma}{\Lambda_3^9}~,
\eeq
where $L_{\mu\alpha\nu\beta}$ is a fairly complicated combination of the Riemann tensor and its contractions, that can be found in \cite{Chkareuli:2011te,deRham:2011by}. Both of these terms (as well as their arbitrary functions) are invariant under the quasidilatations \eqref{newglob} and are as important on the self-accelerated FRW backgrounds, as the rest of the terms considered above; they would be expected to affect the analysis at order one. Moreover, as already emphasized above, these terms lead to the Ostrogradsky ghost-free scalar-tensor interactions of $\sigma$ with the metric perturbation in the Minkowski space decoupling limit, retaining the remarkable non-renormalization properties of the theory.

It is important to stress, that while the very existence of selfaccelerated solutions with the Hubble parameter 
of order of the graviton mass might also be the property of the full quantum theory, the nature of these backgrounds and their
perturbations  seems to crucially depend on an unknown UV extension 
of the theory at and above the $\Lambda_3$ energy scale. Knowledge of a putative 
UV theory with the quasidilaton, valid all the way up to the Planck scale, 
becomes a necessity  for the full understanding of homoegenous and isotropic flat FRW backgrounds in this theory.

It is clear that for the flat FRW solutions at hand, due to the large time variation of the quasidilaton field, $\dot\sigma\sim \Lambda_2^2$,  both the late time ($H\sim m$), as well as the early ($H\gg m$) cosmology are sensitive to the unknown UV dynamics of the theory. This is unlike the pure massive gravity theories, for which all of the previously obtained solutions are well within the effective field theory domain, possible due to the parametric separation of the cosmological and the UV scalaes, $m\ll \Lambda_3$. Now, in the limit $\omega\to\infty$, the $\sigma$ field is expected to decouple and the dynamics of the theory, along with its solutions, should be dominated by the pure massive GR \cite{D'Amico:2012zv}. In particular, for sufficiently large $\omega$, one expects the existence in these theories of the inhomogeneous solutions of massive GR, that recover the standard early cosmology due to the Vainshtein mechanism \cite{D'Amico:2011jj}.

\begin{acknowledgements}
We thank the authors of Ref. \cite {Paper}  for sharing their preliminary 
results and for pointing out a missing term in our computation. G.D'A is supported by the James Arthur Fellowship, GG is supported by NSF  grant PHY-0758032 and NASA grant NNX12AF86G S06, 
LH is supported by DOE DE-FG02-92-ER40699 and NASA NNX10AN14G, and DP has been supported by the U.S. Department of Energy under contract No. DOE-FG03-97ER40546.
\end{acknowledgements}
\appendix
\section{}

The most general ansatz for flat, homogeneous and isotropic solutions is given as follows:
\beq
d s^2=- N^2(t)  d t^2+a^2(t) d \vec x^2 \, , \quad \phi ^0 = f(t) \, , \quad  \phi ^i = x^i \, , \quad \sigma = \sigma(t) ~.
\eeq
One can substitute this into \eqref{eq:einstein} to obtain the minisuperspace action\footnote{Note that we have retained the lapse $N(t)$ in the action despite the fact that by time reparametrization invariance it can be fixed to an arbitrary value as long as $f(t)$ does not equal to one. However, keeping it explicitly is quite convenient, since it allows to quickly derive a first-order Friedmann equation for the scale factor.}. Varying the resulting expression w.r.t. $f$ (that plays the role of a Lagrange multiplier in the action) yields the modified constraint, which can be solved by 
\beq
\label{sols}
e^{\sigma/\mpl}=ca, ~ ~c\(1+\frac{3}{•4}\alpha_3+\alpha_4-\(1+\frac{3}{2}\alpha_3+3\alpha_4\)c+\frac{3}{4} \(\alpha_3+4\alpha_4\)c^2-(\alpha_4+\beta)c^3\)=0
\eeq
The Friedmann equation is obtained by varying with respect to $N$ and subsequently setting $N\to 1$. This yields a de Sitter metric with the Hubble scale $H$ of order of the graviton mass $m$.
\be
\label{eq:frwc}
\begin{split}
H^2 =
~\frac{m^2 \bigg( \frac{1}{4} (\al_3 + 4 \al_4) c^3 
- (1+\frac{3}{2} \al_3 + 3 \al_4) c^2 
+ (3 + \frac{9}{4} \al_3 + 3 \al_4) c
- (2 + \al_3 + \al_4) \bigg) }{1-\frac{\omega}{•6}}~.
\end{split}
\ee
Finally, the Lagrange multiplier $f=\phi^0$ can be obtained from the $\sigma$ equation of motion and is given as follows
\beq
\phi^0=\bar c\int\frac{dt}{a(t)}, \quad \bar c=const=1 + \frac{\om}{\kappa }~ \frac{H^2}{m^2}, 
\eeq
where 
\beq
\kappa = c ~\bigg [3 \(1 + \frac{3}{4} \al_3 + \al_4\)
- 2 \(1+\frac{3}{2} \al_3 + 3 \al_4\) c 
+  \frac{3}{4} \(\al_3 + 4 \al_4\) c^2 \bigg ]~.
\eeq

To study the perturbations, it will be more convenient to work in terms of conformal time $\tau$, transforming to an "almost unitary" gauge in which the background metric is $g_{\mn}=a^2(\tau)\eta_{\mn}$ and the auxiliary scalars are \emph{frozen} to their background values $\phi^0=\bar c\tau$, $\phi^i=x^i$. We define the perturbations of the dynamical fields in this gauge as follows,
\be
g_{\mn}=a^2(\eta_{\mn}+h_{\mn}), \qquad \sigma/\mpl=\ln(ca)+\zeta.
\ee
The tensor $\tK$, up to quadratic order in perturbations is given by the following expression
\begin{align}
\tK^\mu_{~\nu}&=\delta^\mu_\nu-c~a~ e^{\zeta}\sqrt{\frac{1}{a^2}(\eta^{\mu\lambda}-h^{\mu\lambda}+h^{\mu\rho}h_{\rho}^{~\lambda}+\dots) \Sigma_{\lambda\nu}}\nn\\
&=\delta^\mu_\nu-c~ (1+\zeta+\frac{1}{2}\zeta^2+\dots)\sqrt{\Sigma^\mu_{~\nu}-h^{\mu\lambda}\Sigma_{\lambda\nu}+h^{\mu\rho}h_{\rho}^{~\lambda}\Sigma_{\lambda\nu}+\dots) }
\end{align}
where all indices are assumed to be raised/lowered with the flat Minkowski metric and $\Sigma^{\mu}_{~\nu}\equiv\p^\mu\phi^a\p_\nu\phi^b\eta_{ab}=diag(\bar c^2,1,1,1)$. We will also need an expansion to the quadratic order of the metric determinant
\be
\sqrt{-g}=1+\frac{1}{2}h+\frac{1}{8}h^2-\frac{1}{4}h^{\mn}h_{\mn}+\dots~.
\ee
The Einstein frame action we would like to perturb can be conveniently parametrized by separating the pure general relativity sector in the following way 
 \begin{align}
\label{sublambdatilde0}
S_E&= \frac{1}{2}\int \sqrt{-g}~  [ R-6 H^2] \nn \\
&+ \frac{1}{2}\int \sqrt{-g}\left[ 6 H^2 -\frac{\omega}{\mpl^2} g^{\mn}\p_\mu\sigma\p_\nu\sigma
- \frac{m^2}{4} \( \mathcal{U}_2(\tK) +\alpha_3 \mathcal{U}_3(\tK)+\alpha_4 \mathcal{U}_4(\tK) \) \right]\nn\\&+3\mpl ^2 m^2 \int d^4 x~\beta~ e^{4\sigma/\mpl}\sqrt{-det~ \p_\mu\phi^a\p_\nu\phi_a}~.
\end{align}
The first term describes pure GR on dS space (with a CC, consistent with the expansion rate), while the perturbations of the rest of the Lagrangian will describe deviation from GR.
The quadratic perturbations of the second line of \eqref{sublambdatilde0} can be written as follows (note that the indices on metric perturbations are \emph{not} raised and prime denotes a derivative w.r.t. conformal time $\tau$)
\begin{align}
\label{pert_h}
S^{(2)}_E\supset \frac{1}{2}\int d^4x ~a^4\bigg\{  &\frac{\omega}{a^2} \(\zeta'^2-(\p_i \zeta)^2\) +\frac{\omega H}{a} (h_{00}+h_{ii})\zeta' - \frac{2\omega H}{a} h_{0i} \p_i \zeta +(\gamma_1 h_{00}+\gamma_2 h_{ii})\zeta\nn \\
&+\gamma_3 h_{00}^2+\gamma_4h_{00}h_{ii}+ \gamma_5 h_{0i}h_{0i}+\gamma_6 h_{ij}h_{ij}+\gamma_7 h_{ii}^2+\gamma_8\zeta^2\bigg\}~.
\end{align}
Here, with the definitions
\beq
\label{betas}
x_0=4+3\alpha_3+4\alpha_4, \quad 
x_1=2+3\alpha_3+6\alpha_4, \quad 
x_2=\alpha_3+4\alpha_4, \quad 
x_3=2+\alpha_3+\alpha_4 ~,
\eeq
the coefficients appearing in the action are
\begin{align}
\label{gammas}
\gamma_1&=3m^2\kappa\nn\\
\gamma_2&=-\frac{1}{4} m^2 c\(9 \bar c c^2 x_2-4 (1+2\bar c) c x_1+3 (2+\bar c)x_0\)\nn\\
\gamma_3&=\frac{\omega}{4}H^2\nn\\
\gamma_4&=-\frac{m^2}{2}\kappa\\
\gamma_5&=\frac{\kappa}{1+\bar c} m^2 \nn\\
\gamma_6&=\frac{1}{16} m^2 \( 3 \bar c c^3 x_2 -2(1+3\bar c) c^2 x_1 +3(3+2\bar c) c x_0-24 x_3\)-\frac{H^2}{4}(\omega+6)\nn\\
\gamma_7&=\frac{1}{16} m^2 \( 2 \bar c c^2 x_1 -3 (1+\bar c) c x_0+12 x_3\)+\frac{H^2}{8}(\omega+6)\nn \\
\gamma_8&=\frac{3}{4} m^2c\(64 \bar c c^3(\alpha_4+\beta)-9c^2 (1+3\bar c) x_2+8(1+\bar c)c x_1-(3+\bar c) x_0\)\nn~.
\end{align}

\bibliography{bibliography}

\end{document}